\documentclass[article,5p]{elsarticle}
\journal{Chemical Engineering Science}

\usepackage[utf8]{inputenc}
\usepackage[english]{babel}

\usepackage{graphicx}
\usepackage{xcolor}
\usepackage{natbib}
\usepackage{amsmath}
\usepackage{amssymb}
\usepackage{bm}
\usepackage{ar}

\newcommand{\etal}{et al.\ }
\newcommand{\ie}{i.e.,\ }

\newcommand{\secref}[1]{\mbox{section \ref{#1}}}

\newcommand{\tabref}[1]{\mbox{table \ref{#1}}}

\newcommand{\equref}[1]{\mbox{equation (\ref{#1})}}

\newcommand{\figref}[2][]{\mbox{figure \ref{#2}(#1)}}
\newcommand{\figrefSC}[1]{\mbox{Figure \ref{#1}}}
\newcommand{\figrefS}[1]{\mbox{figure \ref{#1}}}

\newcommand{\aver}[1]{\left\langle {#1} \right\rangle}

\begin{document}
\begin{frontmatter}

\title{The impact of porous walls on the rheology of suspensions}

\author[myu1]{Marco E. Rosti\corref{cor1}}
\ead{marco.rosti@oist.jp}
\address[myu1]{Complex Fluids and Flows Unit, Okinawa Institute of Science and Technology Graduate University, 1919-1 Tancha, Onna-son, Okinawa 904-0495, Japan}
\cortext[cor1]{Corresponding author}

\author[myu2]{Parisa Mirbod}
\address[myu2]{Department of Mechanical and Industrial Engineering, The University of Illinois at Chicago, Chicago, USA}

\author[myu3]{Luca Brandt}
\address[myu3]{Linn\'{e} FLOW Centre and SeRC, Department of Engineering Mechanics, KTH Royal Institute of Technology, Stockholm, Sweden}

\begin{abstract}
We study the effect of isotropic porous walls on a plane Couette flow laden with spherical and rigid particles. We perform a parametric study varying the volume fraction between $0$ and $30\%$, the porosity between $0.3$ and $0.9$ and the non-dimensional permeability between $0$ and $7.9 \times 10^{-3}$ We find that the porous walls induce a progressive decrease in the suspension effective viscosity as the wall permeability increases. This behavior is explained by the weakening of the wall-blocking effect and by the appearance of a slip velocity at the interface of the porous medium, which reduces the shear rate in the channel. Therefore, particle rotation and the consequent velocity fluctuations in the two phases are dampened,  leading to reduced particle interactions and particle stresses. Based on our numerical evidence, we provide a closed set of equations for the suspension viscosity, which can be used to estimate the suspension rheology in the presence of porous walls.
\end{abstract}

\end{frontmatter}

\section{Introduction} \label{sec:introduction}
One of the challenges faced in every aspect of a new technology is how to reduce energy loss and inefficiencies by manufacturing advanced and novel devices at low or no cost. When a suspension transport properties are critical, as of interest here, these devices include, but are not limited to, technologies such as extrusion (shallow screw channels) and thin lubricating films. In such systems, proper boundary conditions play a major role in controlling and driving the flow. In this study, we explore the flow of particle suspensions over porous surfaces in a plane Couette flow in order to evaluate the effect of these walls on the particle laden flow behavior. This understanding may contribute to improving the efficiency and operating lifetime of the abovementioned devices.

Particle-laden flows are encountered in various industrial applications, including blood flow, slurry transport, and pharmaceutical industry applications. Slow flow of non-Brownian suspensions has been analytically and experimentally examined in various geometries, the simplest probably being the Couette flow between impermeable walls \cite{leighton_acrivos_1987b, phillips_armstrong_brown_graham_abbott_1992a, acrivos_mauri_fan_1993a, nott_brady_1994a, morris_boulay_1999a, zarraga_hill_leighton-jr_2000a, singh_nott_2003a, sierou_brady_2002a, miller_morris_2006a, yurkovetsky_morris_2008a, deboeuf_gauthier_martin_yurkovetsky_morris_2009a, miller_singh_morris_2009a, yeo_maxey_2010a, guazzelli_morris_2011a, lashgari_picano_breugem_brandt_2014a}. On the other hand, Newtonian fluid flow past porous surfaces also has many important applications such as flow over sediment beds \cite{goharzadeh_khalili_jorgensen_2005a}, over crop canopies and in forests \cite{kruijt_malhi_lloyd_norbre_miranda_pereira_culf_grace_2000a, ghisalberti_nepf_2009a}, in the human body \cite{guo_weinstein_weinbaum_2000a} and over carbon nanotubes \cite{battiato_bandaru_tartakovsky_2010a}. In particular, flow over porous walls is gaining increasing interest due to the possibility of passively controlling the flow and reducing drag in both laminar \cite{mirbod_wu_ahmadi_2017a} and turbulent flows \cite{rosti_brandt_pinelli_2018a}. However, the dynamics and rheological behavior of particles flowing over porous surfaces are qualitatively different from those observed over smooth surfaces due to modifications of the flow and of particle-induced fluid motions by the porous surface, as we will also document here. 

Einstein \cite{einstein_1956a} was the first to show that, the effective viscosity of a dilute suspension (\ie volume fraction $\Phi \rightarrow 0$) of rigid particles in a Newtonian fluid linearly increases with the particle volume fraction $\Phi$, when inertia is negligible. Later on, Batchelor \cite{batchelor_1977a} and Batchelor and Green \cite{batchelor_green_1972a} extended Einstein's study to higher volume fractions and added a second-order term in $\Phi$. In general, there is no analytical relation able to predict the suspension viscosity at higher volume fractions, and empirical fits are instead used. Here, we will adopt the so-called Eilers fit \cite{ferrini_ercolani_de-cindio_nicodemo_nicolais_ranaudo_1979a, zarraga_hill_leighton-jr_2000a, singh_nott_2003a, kulkarni_morris_2008a}. Deviations from the behaviour predicted by this and similar expressions have been found due to inertia \cite{alghalibi_lashgari_brandt_hormozi_2018a} and at very large volume fractions once friction forces become dominant \cite{fall_huang_bertrand_ovarlez_bonn_2008a, seto_mari_morris_denn_2013a}. Herein, we quantitatively characterize the rheological behavior of particles over porous walls across a sheared suspension at semi-dilute concentrations, $\Phi \lesssim 30\%$ and negligible inertia.

Recently, Rosti \etal \cite{rosti_ardekani_brandt_2019a} studied the rheology of a particle suspension in channels with elastic walls and found a shear-thinning behavior of the suspension. This was caused by the particle migration away from the wall towards the channel center due to a lift force \cite{rallabandi_oppenheimer_zion_stone_2018a} generated by the particle induced wall deformation. In the present work, we focus on a different kind of wall-modification, rigid porous walls where the fluid is allowed to penetrate through the porous walls.

In particular, we employ direct numerical simulations (DNSs) to explore the particle motion and interactions over rigid porous surfaces for a plane Couette flow where both surfaces are covered with porous media with known permeability and porosity. The chosen set-up is the one typical of fundamental rheology studies, but the results can be extended to more complex and realistic geometries, such as channel and duct flows or Taylor-Couette flows. Here, we quantify the variations in the suspension stresses and slip velocity in a plane Couette flow due to the existence of porous surfaces. We also study the combined effects of particle volume fraction and wall permeability on the effective viscosity of the suspension.

\sloppy
The present manuscript is organized as follows: in \secref{sec:formulation} we first present the mathematical and numerical formulations used to model the flow; then, in \secref{sec:result} we discuss the results of the simulations in terms of fluid and particle statistics and their variation with the particle volume fraction and with the parameters characterizing the porous media; finally, we collect the main findings in \secref{sec:conclusion} and draw some final conclusions.

\begin{table*}
\begin{minipage}{0.3\textwidth}
  \begin{center}
   $h_p=h/2$, $\varepsilon=0.6$ \\ \vspace{0.1cm}
  \begin{tabular}{c|ccccc}
  \hline \hline
  $\Phi$		&	$0.00$	&	$0.06$	&	$0.12$	&	$0.24$	&	$0.30$	\\
  $\sigma\times10^{3}$	&	$0$	&	$0$	&	$0$	&	$0$	&	$0$	\\ \hline
  $\Phi$		&	$0.00$	&	$0.06$	&	$0.12$	&	$0.24$	&	$0.30$	\\
  $\sigma\times10^{3}$	&	$0.79$	&	$0.79$	&	$0.79$	&	$0.79$	&	$0.79$	\\ \hline
  $\Phi$		&	$0.00$	&	$0.06$	&	$0.12$	&	$0.24$	&	$0.30$	\\
  $\sigma\times10^{3}$	&	$2.5$	&	$2.5$	&	$2.5$	&	$2.5$	&	$2.5$	\\ \hline
  $\Phi$		&	$0.00$	&	$0.06$	&	$0.12$	&	$0.24$	&	$0.30$	\\
  $\sigma\times10^{3}$	&	$7.9$	&	$7.9$	&	$7.9$	&	$7.9$	&	$7.9$	\\ \hline \hline
  \end{tabular}
  \end{center}
\end{minipage}
\hspace{2cm}
\begin{minipage}{0.3\textwidth}
  \begin{center}
   $h_p=h/2$, $\sigma=7.0 \times 10^{-3}$ \\ \vspace{0.1cm}
  \begin{tabular}{c|ccccc}
  \hline \hline
  $\Phi$		&	$0.00$	&	$0.12$	\\
  $\varepsilon$	&	$0.3$	&	$0.3$	\\ \hline
  $\Phi$		&	$0.00$	&	$0.12$	\\
  $\varepsilon$	&	$0.6$	&	$0.6$	\\ \hline
  $\Phi$		&	$0.00$	&	$0.12$	\\
  $\varepsilon$	&	$0.9$	&	$0.9$	\\ \hline \hline
  \end{tabular}
  \end{center}
\end{minipage}
\begin{minipage}{0.3\textwidth}
  \begin{center}
   $\varepsilon=0.6$, $\sigma=7.0 \times 10^{-3}$ \\ \vspace{0.1cm}
  \begin{tabular}{c|ccccc}
  \hline \hline
  $\Phi$		&	$0.00$	&	$0.12$	\\
  $h_p/h$	&	$0.25$	&	$0.25$	\\ \hline
  $\Phi$		&	$0.00$	&	$0.12$	\\
  $h_p/h$	&	$0.5$	&	$0.5$	\\ \hline
  $\Phi$		&	$0.00$	&	$0.12$	\\
  $h_p/h$	&	$1$	&	$1$	\\ \hline \hline
  \end{tabular}
  \end{center}
\end{minipage}
  \begin{center}
  \caption{Summary of the simulations performed at different particle volume fractions $\Phi$, porosities $\varepsilon$, permeabilities $\sigma$ and porous layer thicknesses $h_p$, all at a fixed Reynolds number $Re=0.1$, for which inertial effects are considered negligible.}
  \label{tab:cases}
  \end{center}
\end{table*}

\section{Methodology} \label{sec:formulation}
\begin{figure}
  \centering
  \includegraphics[width=0.49\textwidth]{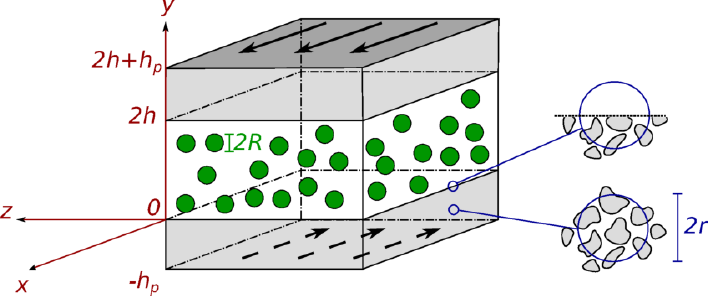}
  \caption{Sketch of the computational domain, the coordinate systems and the different scales involved in the problem.}
  \label{fig:sketch}
\end{figure}
We study the Couette flow of a Newtonian fluid laden with a suspension of rigid particles bounded by two homogeneous and isotropic porous walls. The fluid is incompressible and two flat, isotropic and homogeneous porous layers are attached to the impermeable moving walls, as shown in \figrefS{fig:sketch}. The streamwise, wall-normal and spanwise coordinates are denoted by $x$, $y$ and $z$ ($x_1$, $x_2$, and $x_3$), and similarly $u$, $v$ and $w$ ($u_1$, $u_2$, and $u_3$) are the corresponding velocity components. $y = 0$ and $y=2h$ denote the two interfaces between the porous layers and the fluid region, while $y = -h_p$ and $y= 2h + h_p$ are the location of the bounding impermeable walls, being $h_p$ the porous layer thickness. Rigid spheres, with the same mass density as the carrier fluid and radius $R$, are suspended in the purely fluid region between the two porous slabs.

The flow is governed by the Navier-Stokes equations, with the conservation of momentum and the incompressibility constraint written as
\begin{align}
\label{eq:NS}
\rho \left( \frac{\partial u_i}{\partial t} + \frac{\partial u_i u_j}{\partial x_j} \right) = \frac{\partial \tau_{ij}}{\partial x_j} \;\;\; \textrm{and} \;\;\; \frac{\partial u_i}{\partial x_i} = 0.
\end{align}
\sloppy
In the above, $\rho$ indicates the fluid density and $\tau_{ij}$ the Cauchy stress tensor. The fluid is assumed to be Newtonian with constitutive equations $\tau_{ij}^f = -p \delta_{ij} + 2 \mu \mathcal{D}_{ij}$, where $p$ is the pressure, $\mu$ is the fluid dynamic viscosity, $\mathcal{D}_{ij}$ is the strain rate tensor defined as $\mathcal{D}_{ij}=\left( \partial u_i/\partial x_j + \partial u_j/\partial x_i \right)/2$ and $\delta$ is the Kronecker delta. 
The particle velocity and rotation are governed by the Newton-Euler equations, which can be stated as
\begin{subequations} \label{eq:newton_euler}
\begin{align}
  \rho^p V^p \frac{dU_i^{p_c}}{dt} &= \oint_{\partial V^p} \tau_{ij} n_j dA + F_i^c, \\
  I^p \frac{d \Omega_i^{p_c} }{dt} &= \oint_{\partial V^p} \epsilon_{ijk} r_j \tau_{kl} n_l dA + T_i^c,
\end{align}
\end{subequations}
where $\rho^p$, $I^p$ and $V^p$ are the density, moment of inertia and volume of the particle: when the particle is a rigid sphere with radius $R^p$, we have $V^p=\left(4/3\right) \pi {R^p}^3$ and $I^p = \left(2/5\right) \rho^p V^p {R^p}^2$. In the previous equations, $\tau_{ij}$ is the fluid stress tensor in \equref{eq:NS}, $n_i$ is the unit normal vector pointing outwards from a particle and $F_i^c$ and $T_i^c$ are the particle-particle and particle-wall interaction force and torque. These include a lubrication correction and a soft collision model \citep{costa_boersma_westerweel_breugem_2015a}. In particular, we use Brenner's asymptotic solution \citep{brenner_1961a} to correct the lubrication force when the distance between solid objects is less than a certain threshold and cannot be accurately resolved by the numerical mesh; surface roughness is accounted for by saturating this force at very small distances;  finally, when spheres are in contact, both the normal and tangential contact force components are obtained from the overlap and the relative velocity. We use an immersed boundary method (IBM) to describe the presence of the rigid particles by adding to the right-hand side of the momentum equation a body force $f_i$ that forces the fluid velocity on the particle surface to match the particle velocity \cite{breugem_2012a, izbassarov_rosti_niazi-ardekani_sarabian_hormozi_brandt_tammisola_2018a}. 

We characterize the porous layer by the porosity $\varepsilon$, the volume of void regions divided by the total volume, and the permeability $\mathcal{K}_{ij}$, a tensor measuring the ease to flow through the medium. When the porous medium is isotropic, the permeability can be described by a single scalar quantity $\mathcal{K}$. As already mentioned, the flow through a porous medium is governed by the Navier-Stokes equations with the no-slip boundary conditions imposed on all the porous elements. However, due to the highly complex solid matrix shape and the related resolution requirements, this approach is impractical (except for very simplified cases \citep{de-vita_rosti_izbassarov_duffo_tammisola_hormozi_brandt_2018a}). To overcome these difficulties, it has been proposed  \cite{whitaker_1969a, whitaker_1986a, whitaker_1996a} to model only the large-scale behavior of the flow in the porous medium, averaging (over a small sphere of radius $r$ and volume $V$) the Navier-Stokes equations, as illustrated on the right in figure \ref{fig:sketch}. This procedure leads to the volume-averaged Navier-Stokes equations. Rosti et al. \cite{rosti_cortelezzi_quadrio_2015a} describe a specific form of the VANS equations obtained assuming an isotropic porous medium with negligible fluid inertia and large scale separation (\ie $\ell \ll r \ll L$, with  $\ell$ being the smallest scale of the flow and the porous matrix and $L \sim h$ the scale of the porous layer). The volume-averaged Navier-Stokes equations read 
\begin{equation} 
\label{eq:VANS}
\rho \dfrac{\partial \aver{u_i}^s}{\partial t} = - \varepsilon  \frac{\partial \aver{p}^f}{\partial x_i} + \mu \frac{\partial^2 \aver{u_i}^s}{\partial x_j \partial x_j} - \frac{\mu \varepsilon}{\mathcal{K}} \aver{\widetilde{u}_i}^s \;\;\; \textrm
{,}
 \;\;\; \frac{\partial \aver{u_i}^s}{\partial x_i} = 0,
\end{equation}
where $\widetilde{u}_i$ is the difference between the flow and the porous medium velocity, which is set equal to the wall velocity. The previous equations are obtained by introducing two average operators: the superficial volume average $\aver{\phi}^s = 1/V \int_{V_f} \psi dV_f$, and the intrinsic volume average $\aver{\psi}^f = 1/V_f \int_{V_f} \psi dV_f$ (here $\psi$ is any fluid variable). Note that, the two operators are linearly related by the condition $\aver{\psi}^s = V_f/V \aver{\psi}^f = \varepsilon \aver{\psi}^f$. The superficial and intrinsic volume averages are commonly chosen for the velocity and pressure field, respectively, as discussed in Refs.\ \citep{quintard_whitaker_1994b, whitaker_1996a}. Recently, Kang and Mirbod \cite{kang_mirbod_2019a} examined the porosity effect on a flow using the VANS equations and a transport equation for the kinetic energy. 
 
Our numerical simulations are based on a 3D solver that adopts an IBM for the particles in the purely fluid region $0<y<2h$, while the volume-averaged Navier-Stokes equations (\ref{eq:VANS}) are solved in the two porous layers $-h_p < y < 0$ and $2h < y < 2h+h_p$. This implies that the particle radius is much larger than the pore size. 
Formally, the fluid flow equations are closed by imposing no-slip boundary conditions on the rigid walls, on the rigid porous material surface and on the moving particles.
In the VANS, however, we only need to impose no-slip at the limiting impermeable walls and  proper conditions on the velocity and stresses at the porous-fluid interface located at $y=0$ and $y=2h$. In our formulation, pressure and velocity continuity are enforced at the interface, while the shear stress may display a jump  \citep{ochoa-tapia_whitaker_1995a}, with a magnitude controlled by a parameter $\tau$ that measures the transfer of stress at the porous/fluid interface \citep{minale_2014a,minale_2014b} and that depends on the porous material considered and by the texture of the solid interface \citep{ochoa-tapia_whitaker_1998a}. In this work, we assume $\tau=0$, which guarantees the validity of the interface condition \citep{minale_2014a,minale_2014b}, as experimentally verified by Carotenuto \etal \cite{carotenuto_vananroye_vermant_minale_2015a}. Using these assumptions, the momentum-transfer conditions \citep{ochoa-tapia_whitaker_1995a} at the interfaces ($y=0$ and $y=2h$) can be simplified as
\begin{equation}
\label{eq:OTW_cond}
u_i = \aver{u_i}^s, \;\;\; p = \aver{p}^f, \;\;\; \frac{\partial u}{\partial y} = \frac{1}{\varepsilon} \frac{\partial \aver{u}^s}{\partial y}, \;\;\; \frac{\partial w}{\partial y} = \frac{1}{\varepsilon} \frac{\partial \aver{w}^s}{\partial y}.
\end{equation}
Note that for simplicity, we drop $\aver{\cdot}$ in the notation.

Numerically, we advance the system of equations with an explicit fractional-step method \citep{kim_moin_1985a}, based on the third-order Runge-Kutta scheme. Spatial derivatives are approximated with the second-order central finite-difference scheme on a staggered grid arrangement. The baseline code used in the present work has been extensively validated in the past for laminar and turbulent multiphase problems; the interested reader is referred to Refs.\ \cite{lashgari_picano_breugem_brandt_2016a, rosti_brandt_2017a, izbassarov_rosti_niazi-ardekani_sarabian_hormozi_brandt_tammisola_2018a, rosti_ardekani_brandt_2019a, zade_fornari_lundell_brandt_2019a, rosti_brandt_2020a} for more details on the numerical scheme and for the validation campaign.

\begin{figure}[t]
  \centering
  \includegraphics[width=0.45\textwidth]{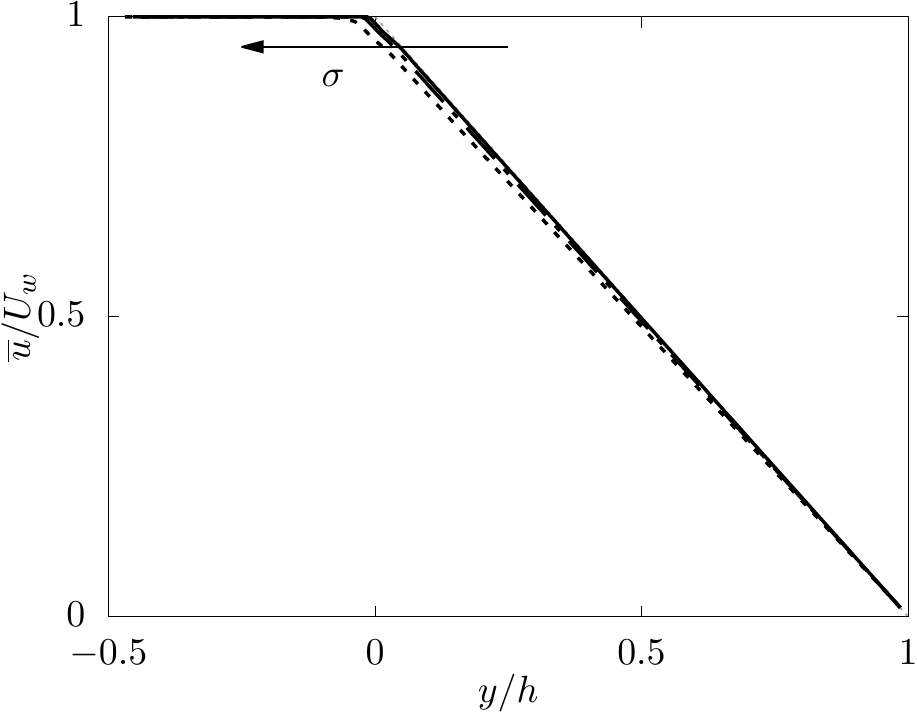}
  \includegraphics[width=0.45\textwidth]{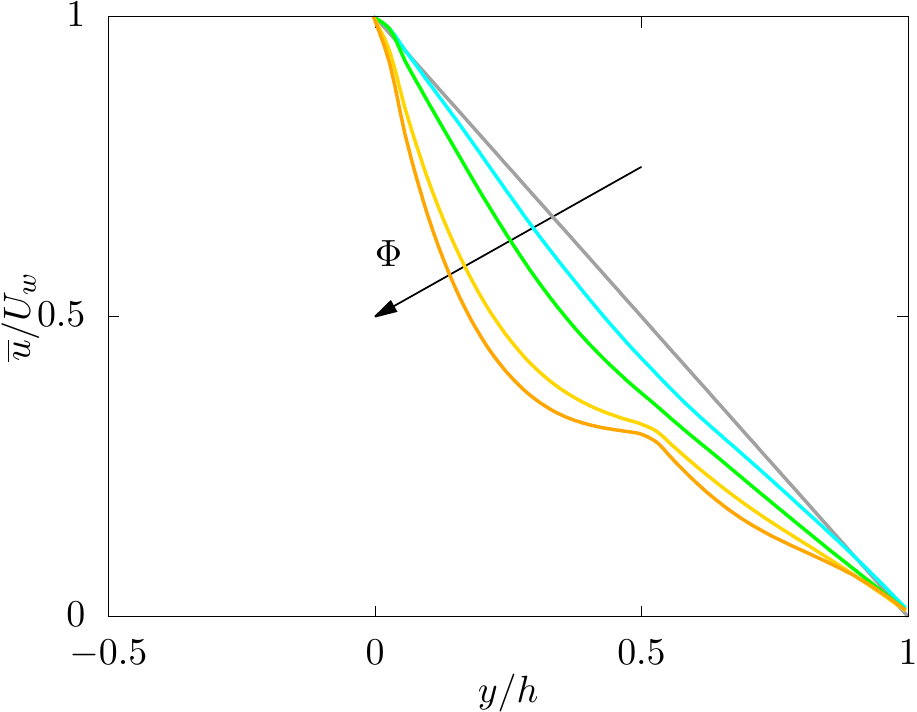}
  \caption{Mean fluid streamwise velocity component $\overline{u}$ as a function of the wall-normal distance $y$ for (top) various wall permeabilities $\sigma$ without any particles, \ie $\Phi=0$, and for (bottom) various particle volume fractions $\Phi$ with impermeable walls, \ie $\sigma=0.0$. In the top panel, the solid, dashed-dotted and dotted line styles pertain to the cases with $\sigma=0.79 \times 10^{-3}$, $2.5 \times 10^{-3}$ and $7.9 \times 10^{-3}$, respectively, while in the bottom panel gray, cyan, green, gold and orange are used to indicate $\Phi=0$, $0.06$, $0.12$, $0.24$ and $0.3$, respectively.}
   \label{fig:vel_ref}
\end{figure}
\begin{figure}[t]
\centering
\includegraphics[width=0.45\textwidth]{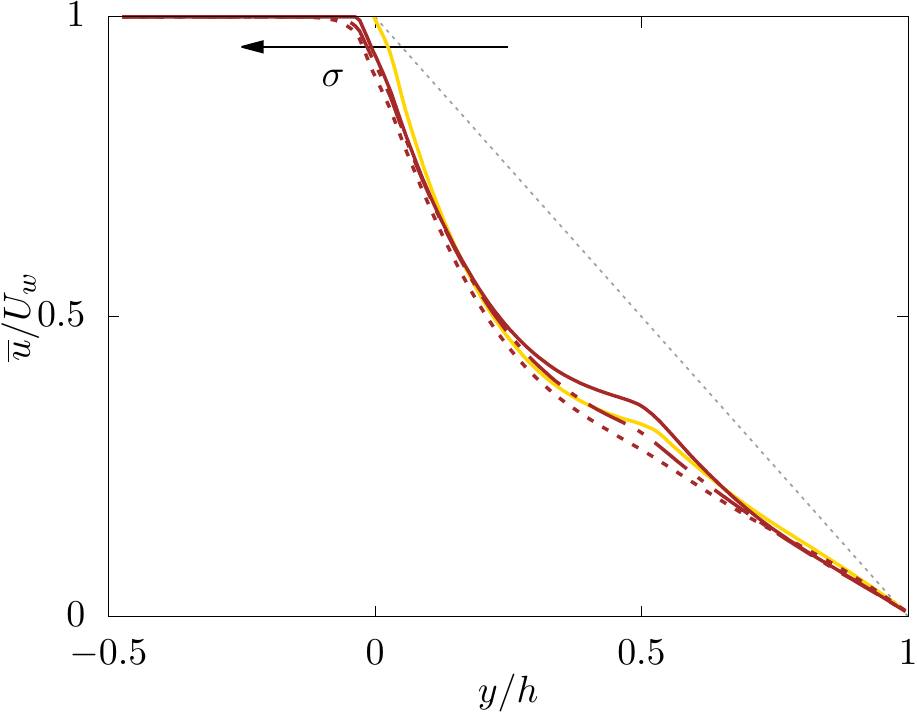}
\includegraphics[width=0.45\textwidth]{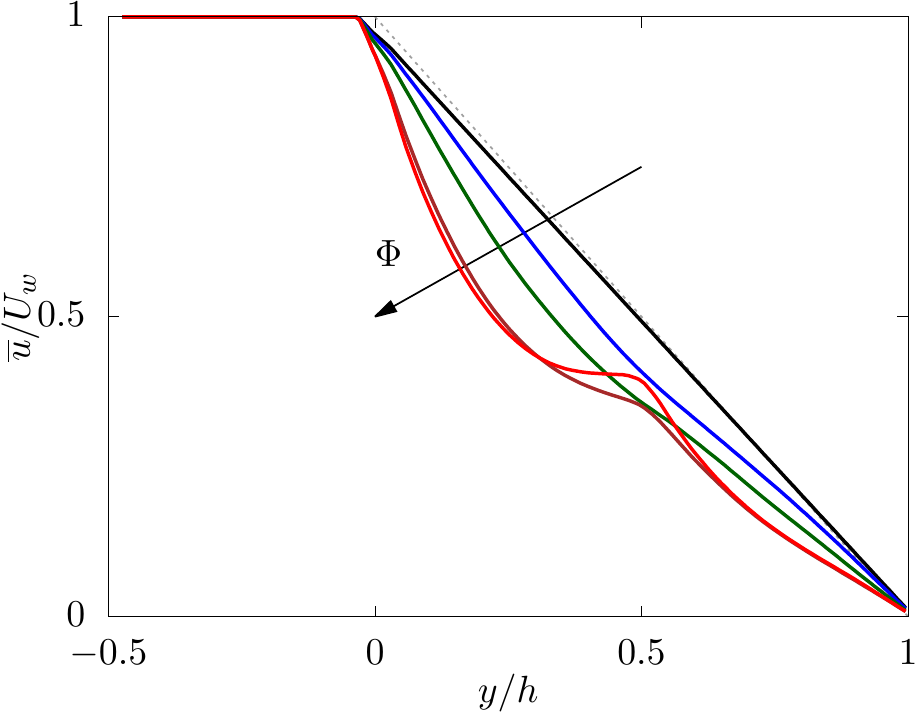}
\caption{Mean fluid streamwise velocity component $\overline{u}$ as a function of the wall-normal distance $y$ for (top) various wall permeabilities $\sigma$ at a fixed particle volume fraction $\Phi=0.24$ and (bottom) various particle volume fractions $\Phi$ with $\sigma=0.79 \times 10^{-3}$. The black, blue, green, brown and red colors are used to distinguish $\Phi=0$, $0.06$, $0.12$, $0.24$ and $0.3$, while the solid, dashed-dotted and dotted line styles are used to distinguish $\sigma=0.79 \times 10^{-3}$, $2.5 \times 10^{-3}$ and $7.9 \times 10^{-3}$, respectively. In the top figure, the gold solid line represents the reference solution over impermeable walls with particles at $\Phi=0.24$.}
\label{fig:vel}
\end{figure}
\begin{figure*}[t]
  \centering
  \includegraphics[width=0.45\textwidth]{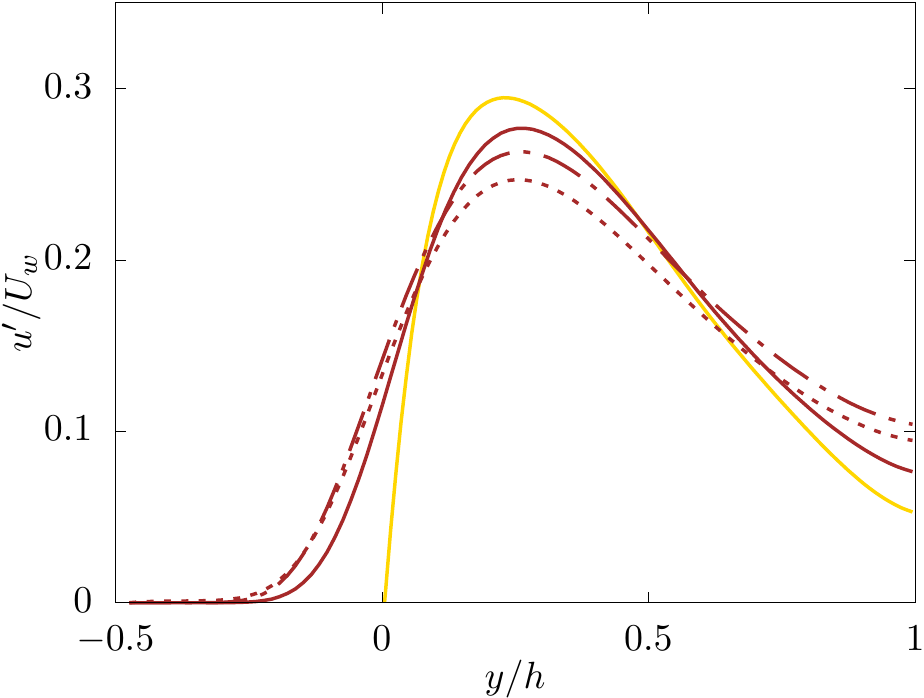}
  \includegraphics[width=0.45\textwidth]{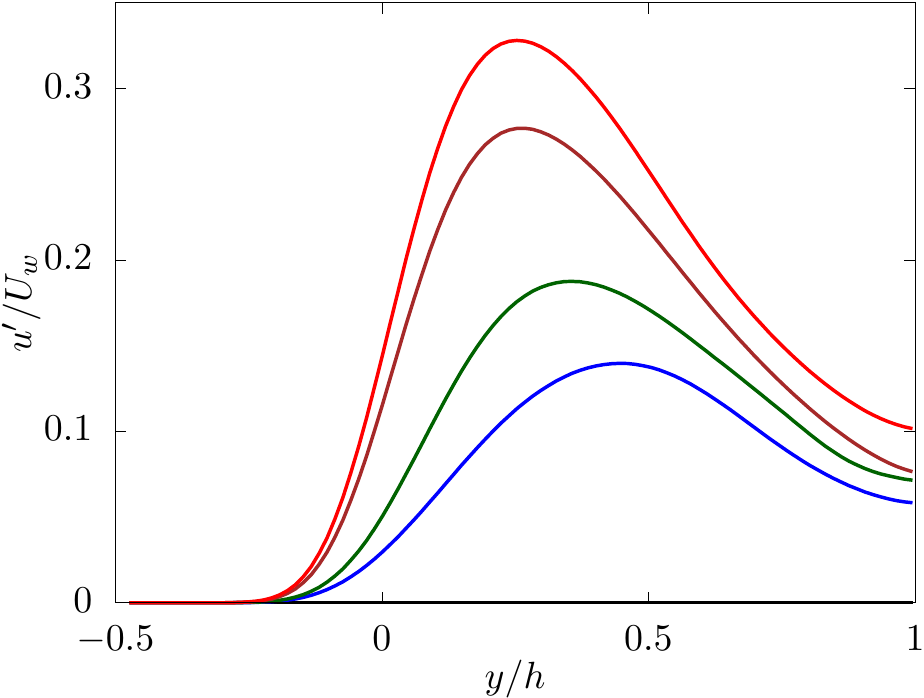}
  \includegraphics[width=0.45\textwidth]{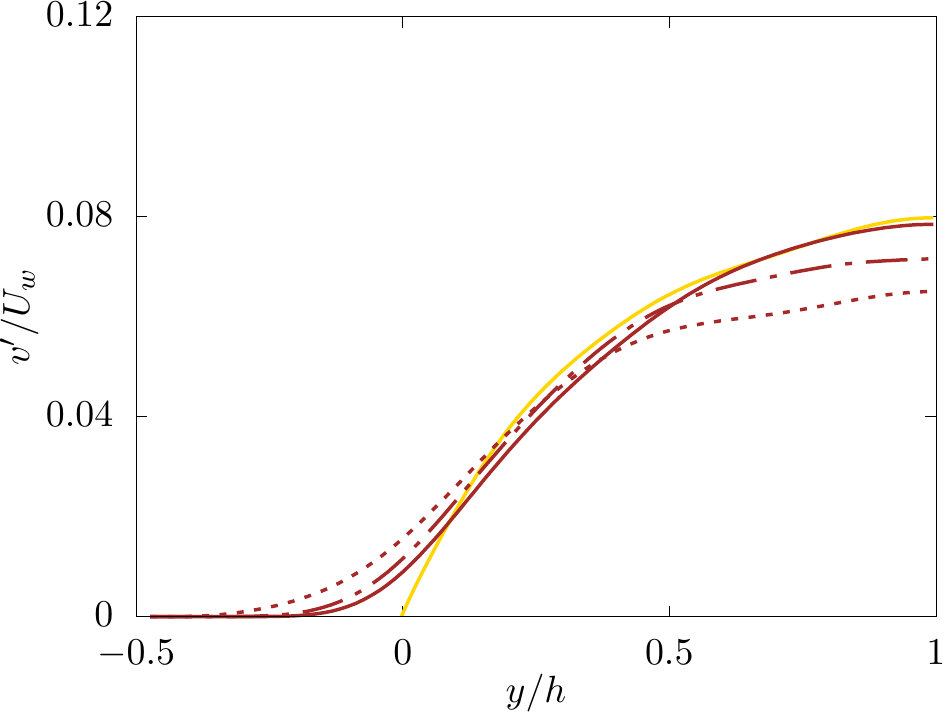}
  \includegraphics[width=0.45\textwidth]{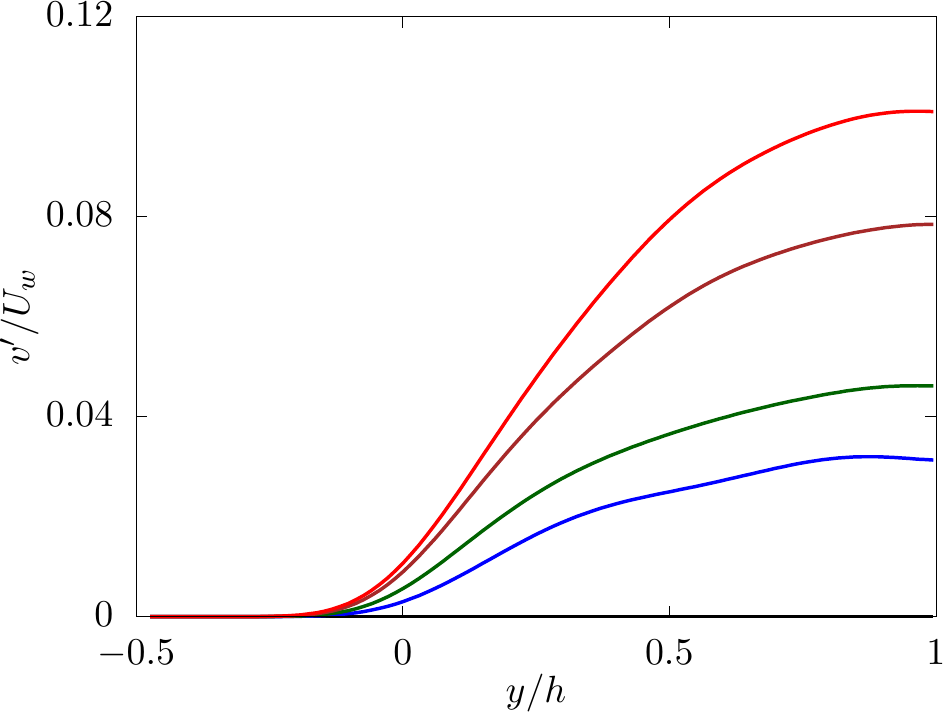}
  \caption{(top) Streamwise and (bottom) wall-normal fluid velocity fluctuations $u'$ and $v'$ as a function of the wall-normal distance $y$ for (left) various wall permeabilities $\sigma$ at a fixed particle volume fraction $\Phi=0.24$ and (right) various particle volume fractions $\Phi$ at fixed permeability $\sigma=0.79 \times 10^{-3}$. In the figure gold solid lines represent the reference solution over impermeable walls with particle volume fraction $\Phi=0.24$. The black, blue, green, brown and red colors are used to distinguish $\Phi=0$, $0.06$, $0.12$, $0.24$ and $0.3$, while the solid, dashed-dotted and dotted line styles are used to distinguish $\sigma=0.79 \times 10^{-3}$, $2.5 \times 10^{-3}$ and $7.9 \times 10^{-3}$, respectively. In the left figures, the gold solid line represents the reference solution over impermeable walls with particles at $\Phi=0.24$.}
  \label{fig:flu}
\end{figure*}
\subsection{Setup}
We consider the Couette flow of a Newtonian fluid laden with rigid spherical particles with radius $R=h/5$. The two rigid walls move with opposite velocity $U_w$ such that the Reynolds number of the simulation is fixed to $Re = \rho \dot{\gamma} R^2/\mu = 0.1$, where $\dot{\gamma}=2U_w/2h$ is the reference shear rate; therefore, we can consider inertial effects to be negligible. Two porous layers of thickness $h_p$ and porosity $\varepsilon$ move with the wall velocity, and the purely fluid region of the domain is bounded by these layers. The numerical domain has size $16R \times 10R+2h_p \times 16R$ and periodic boundary conditions are enforced in the streamwise $x$ and spanwise $z$ directions. We consider three values of the  nondimensional permeability $\sigma=\sqrt{K}/h$, covering the  range $\sigma \in \left[ 0.79, 7.9 \right] \times 10^{-3}$, all in the small permeability limit due to the hypothesis of negligible inertia inside the porous layers;  the perfectly impermeable case,  $\sigma=0$, is used as a reference. In most of our simulations, we fix the porous layer thickness to $h_p=h/2$ and the porosity to $\varepsilon=0.6$; however, we also evaluate the effect of these parameters by simulating selected cases with $h_p=h/4$ and $h_p=h$ and $\varepsilon=0.3$ and $0.9$. Rigid spherical particles are suspended in the purely fluid region of the domain and their volume fraction $\Phi$ is varied in the range $\Phi \in \left[ 0:0.3 \right]$, corresponding to $N_p$ particles suspended in the fluid domain, in particular  $N_p=183$ at $\Phi=0.3$. Note that the particle volume fraction is computed taking only the purely fluid region into account. The full set of simulations analyzed in this work is reported in \tabref{tab:cases}. These parameters are chosen  to facilitate comparisons with previous studies in literature \citep{rosti_brandt_mitra_2018a, rosti_brandt_2018a, alghalibi_lashgari_brandt_hormozi_2018a, rosti_de-vita_brandt_2019a}. In all the simulations, the numerical domain is discretized on a Cartesian uniform mesh with $32$ grid points per sphere diameter $2R$. Finally, note that the size of the domain in the periodic and wall-normal directions is large enough to ensure the independence of the macroscopic suspension properties  from these parameters \citep{fornari_brandt_chaudhuri_lopez_mitra_picano_2016a}. For the present configurations, we have verified that an increase in the domain size by $50\%$ in each direction, results in a change in the mean effective viscosity (computed at the statistically steady state dynamics) lower than $3\%$. Initially, 
particles are positioned randomly in the domain, and the fluid and particles are at rest. When the two walls start moving, the initial transient occurs, and after approximately $60\dot{\gamma}^{-1}$, a statistically steady condition is reached. We computed the mean quantities by averaging over a time of $40 \dot{\gamma}^{-1}$ after this initial transient.

\section{Results} \label{sec:result}
We start our analysis by showing in \figref[top]{fig:vel_ref} the profile of the streamwise component of the mean velocity $\overline{u}$ in the absence of particles, where the overbar indicates the average over the homogeneous directions, i.e.\, $x$ and $z$, and over time. We observe that in the purely fluid region $y>0$, the velocity is linear but with a smaller slope than the nominal one of the flow over impermeable walls $\dot{\gamma}=U_w/h$. The change in the slope grows with the wall permeability $\sigma$, which is due to the weakening of the no-slip boundary conditions at the fluid-porous interface $y=0$, where the difference between the wall velocity $U_w$ and the mean flow assumes a nonzero value, the so-called slip velocity $U_s = \Delta \overline{u}_s = U_w - \overline{u} \left( y=0 \right)$. Inside the porous layers $y<0$, the velocity gradient rapidly decreases, and the velocity eventually reaches the wall velocity $U_w$. The small values of permeability, used here to satisfy the hypothesis of negligible inertia inside the porous layers, induce a slip velocity that is rather limited; however, the consequent change in shear rate is not negligible (up to $7\%$ the nominal value). This is because the permeability has an impact across the whole domain and a strong effect on the particle suspensions as we discuss next.

When particles are suspended in the fluid, the mean velocity profile is modified by their presence, usually with a significant increase of the wall shear stress. This is shown in \figref[bottom]{fig:vel_ref}, where the streamwise mean velocity profile $\overline{u}$ is reported for different volume fractions $\Phi$ in the case of impermeable walls. We observe that the deviation from the linear velocity profile grows with the volume fraction $\Phi$ and results in a fluid velocity lower than that obtained without the particles. Additionally, note that the shear rate across the domain is not uniform in the presence of particles. This displays the tendency of the particles to form preferential layers.

\begin{figure}[t]
  \centering
  \includegraphics[width=0.45\textwidth]{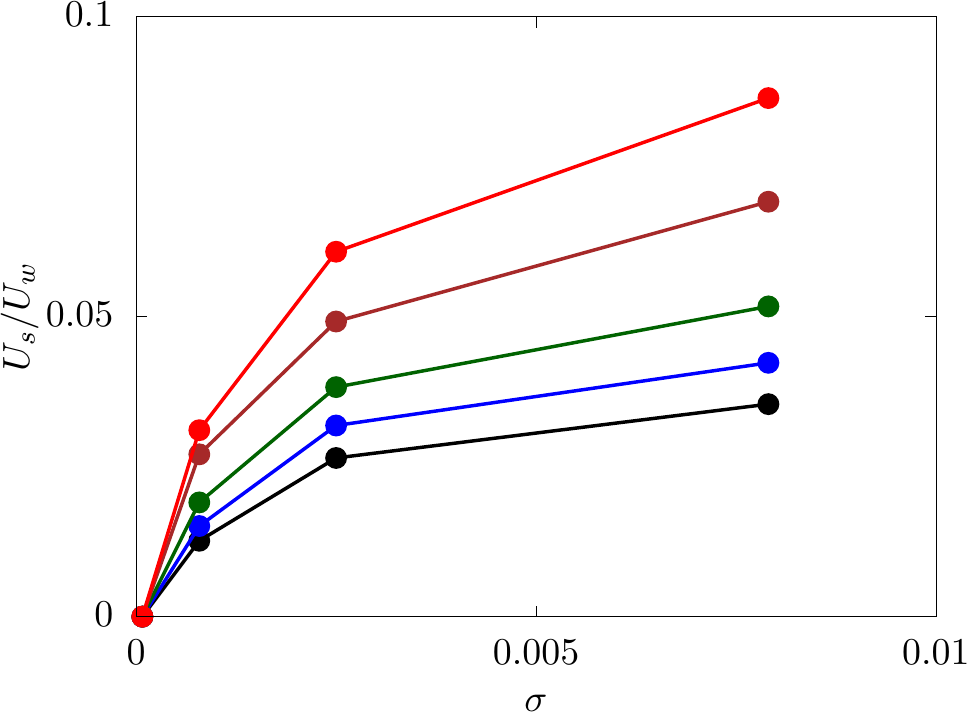}
  \includegraphics[width=0.45\textwidth]{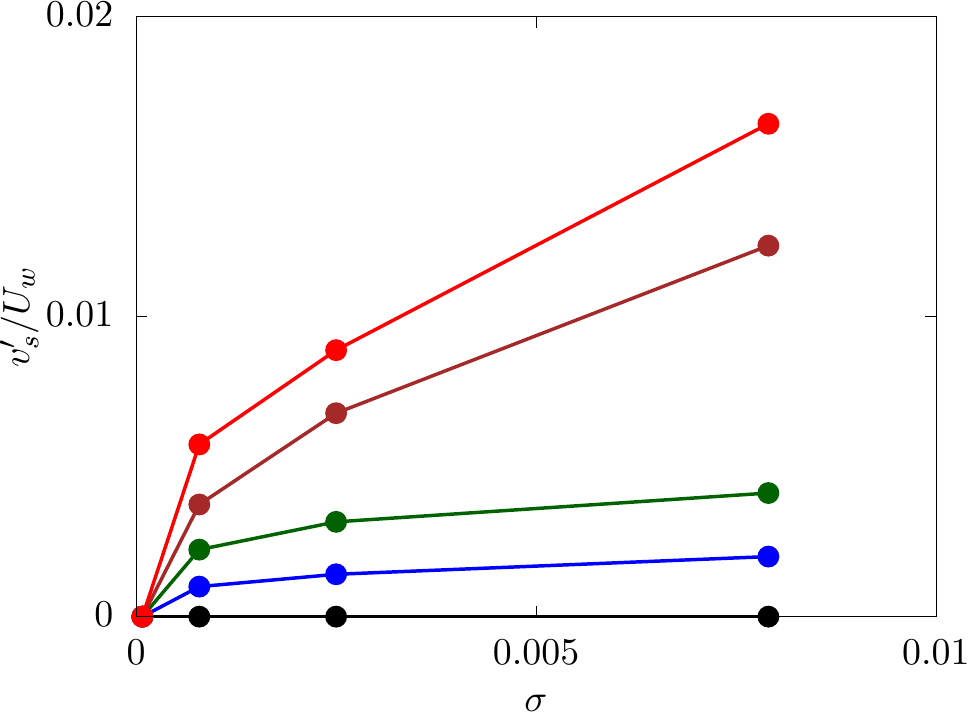}
  \caption{(top) Interface slip velocity $U_s$ and (bottom) wall-normal fluctuations $v'_s$ as a function of the wall permeability $\sigma$ for various particle volume fractions $\Phi$. The black, blue, green, brown and red colors are used to distinguish $\Phi=0$, $0.06$, $0.12$, $0.24$ and $0.3$.}
  \label{fig:wall}
\end{figure}
The combined effect of the permeable walls and the presence of the rigid particles is displayed in \figrefS{fig:vel}, where the top figure shows the modifications of the mean velocity profile due to the wall permeability $\sigma$ at fixed volume fraction $\Phi=0.24$, whereas the bottom panel depicts the modifications in the case of $\sigma=0.70 \times 10^{-3}$ for different values of the solid volume fraction $\Phi$. From the results reported in both figures, we can observe that the two effects combine in a nontrivial way; in particular, we note that both the wall permeability and the addition of particles enhance the magnitude of the slip velocity: the former induces velocity deviations that penetrate deeper in the porous layers, whereas the latter does not, and thus, the increased slip velocity is accompanied by an increased shear rate at the interface. Furthermore, it can be seen that the wall permeability smoothens the velocity profile  in the bulk of the fluid in the presence of particles.

\begin{figure}[t]
  \centering
  \includegraphics[width=0.45\textwidth]{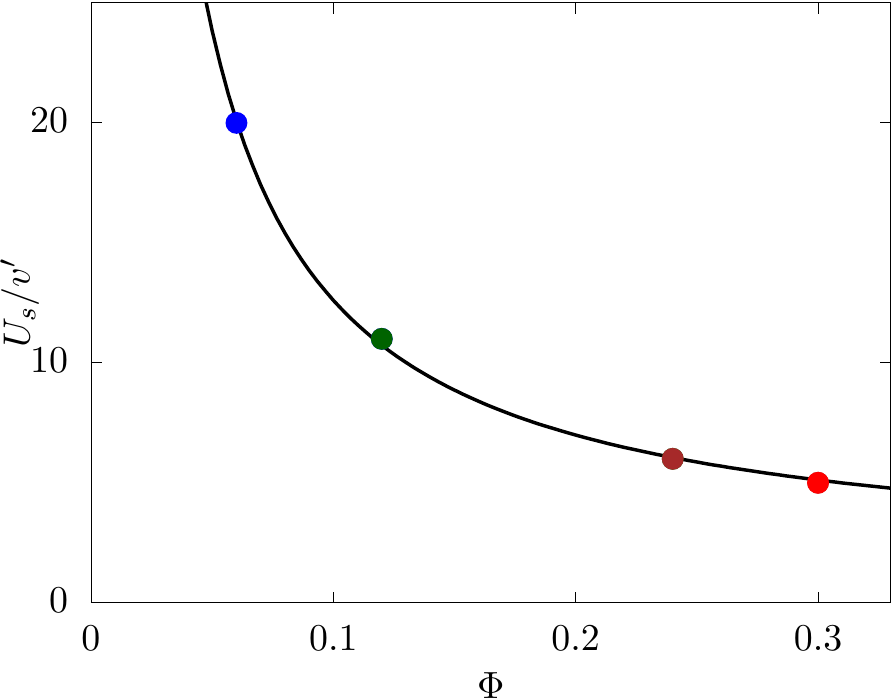}
  \caption{Ratio between the interface slip velocity $U_s$ and the wall-normal fluctuations $v'_s$ averaged for different $\sigma$ as a function of the particle volume fraction $\Phi$. The blue, green, brown and red symbols are used to distinguish $\Phi=0$, $0.06$, $0.12$, $0.24$ and $0.3$. The black solid line is a fit to our data in the form $U_s/v'_s = 1.37 + 1.12 \Phi^{-1}$.}
  \label{fig:wall_ratio}
\end{figure}
In the absence of particles, the flow is stationary, and the velocity fluctuations are zero, independent of the level of permeability of the porous wall. On the other hand, when particles are suspended in the flow, their motion induces velocity fluctuations, and the flow becomes unsteady. The root-mean square of the fluid streamwise and wall-normal velocity fluctuations $u'$ and $v'$ are reported in \figrefS{fig:flu} for different levels of wall permeability $\sigma$ and particle volume faction $\Phi$. The results show that the streamwise velocity fluctuations are larger than their wall-normal counterparts in all the cases studied here. 

\begin{figure}[t]
  \centering
  \includegraphics[width=0.45\textwidth]{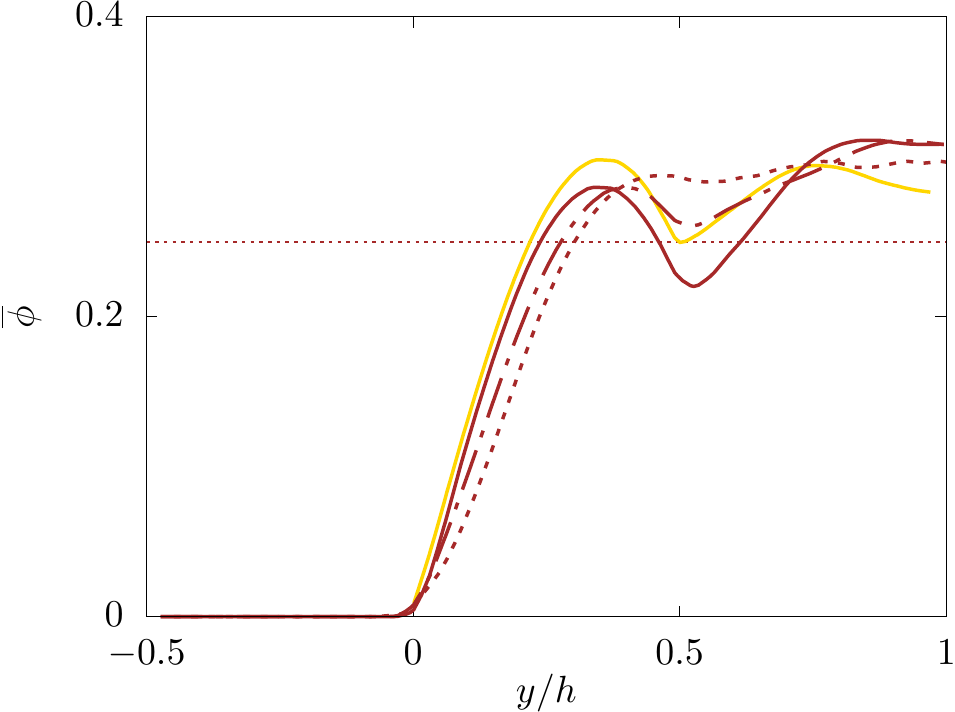}
  \includegraphics[width=0.45\textwidth]{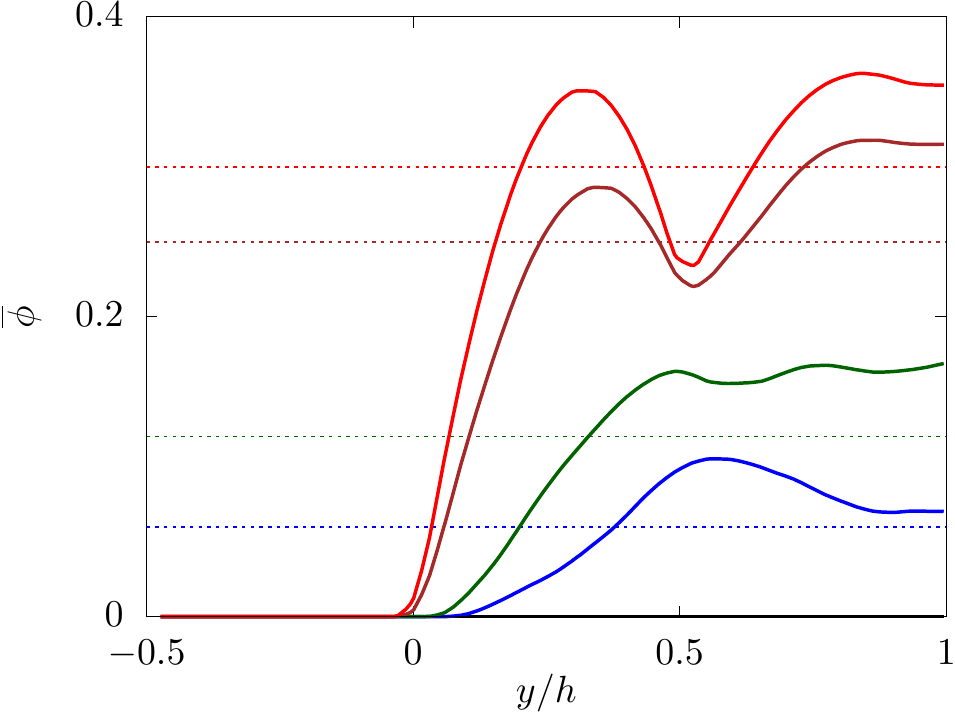}
  \caption{Average particle concentration $\overline{\phi}$ as a function of the wall-normal distance $y$ for (top) various wall permeabilities $\sigma$ at a fixed particle volume fraction $\Phi=0.24$ and (bottom) various particle volume fractions $\Phi$ at fixed permeability $\sigma=0.79 \times 10^{-3}$. In the figure, the horizontal dotted lines represent the bulk particle concentration $\Phi$. The black, blue, green, brown and red colors are used to distinguish $\Phi=0$, $0.06$, $0.12$, $0.24$ and $0.3$, while the solid, dashed-dotted and dotted line styles are used to distinguish $\sigma=0.79 \times 10^{-3}$, $2.5 \times 10^{-3}$ and $7.9 \times 10^{-3}$, respectively. In the top figure, the gold solid line represents the reference solution over impermeable walls with particles at $\Phi=0.24$.}
  \label{fig:phi}
\end{figure}
The wall permeability reduces the peak of both fluctuation components but enhances the velocity fluctuations at the interface due to a reduced wall-blocking effect, similar to what was observed for the mean velocity profile and the rise of the slip velocity. However, velocity fluctuations can penetrate deeper in the porous layers, almost reaching the bounding impermeable walls for the largest permeability shown here. Also, similar to what was observed for the mean velocity profile, the particle volume fraction does not significantly influence the level of penetration of the velocity fluctuations into the porous layers, but enhances the values at the interface and across the whole domain. Thus, in general, while the presence of particles induces velocity fluctuations that increase with their volume fraction, the wall permeability overall reduces the level of fluctuations into the bulk of the domain by dissipating them in the porous layer. Interestingly, 
 the streamwise velocity fluctuations are enhanced by the wall permeability in the center of the channel, whereas the wall-normal component is reduced. Considering that the particles are the origin of these fluctuations, this result suggests that  the particle dynamics are modified in the presence of permeable walls, as analysed below.

\figrefSC{fig:wall} reports the values of the slip velocity $U_s$ and of the wall-normal velocity fluctuations $v'_s$ at the interface $y=0$  ($v'_s=v' \left( y=0 \right)$) for all the cases studied in the present work. The figures confirm that both quantities increase with increasing permeability of the walls, \ie $\sigma$, and with particle volume fraction $\Phi$. In addition, while the slip velocity is not null in the absence of particles, the velocity fluctuations are present only in the particle-laden flows. Both quantities rapidly grow from $0$ even for small values of $\sigma$ and appear to almost saturate for the highest values of $\sigma$ considered, especially for the lower volume fractions investigated in this work. Interestingly, the ratio between the slip velocity of the mean flow $U_s$ and the wall-normal velocity fluctuations $v'_s$  at the interface is approximately independent of wall permeability $\sigma$ but strongly reduces with the particle volume fraction $\Phi$; in particular, the ratio $U_s/v'_s$ goes to infinity for $\Phi=0$ (since the velocity fluctuations are null) and decreases for large volume fractions; see \figrefS{fig:wall_ratio}. We can thus define a function $\mathcal{F}$ that depends only on $\Phi$ such that $U_s=v'_s \mathcal{F} \left( \Phi \right)$; a fit to our data provides the form $\mathcal{F} \left( \Phi \right) = 1.37 + 1.12 \Phi^{-1}$, which is also reported in \figrefS{fig:wall_ratio} and may be useful for future modeling works.

The modifications of the flow that arise when particles are suspended in a channel with permeable walls also affect the particles dynamics. First, we examine the mean local particle concentration $\overline{\phi}$ across the channel. \figrefSC{fig:phi} reports the averaged results for various particle volume fractions and wall permeabilities. We observe that similar to what is typically observed for flows over rigid walls, when the particle volume fraction is increased, the particle concentration increases nonuniformly across the domain and particle layers form, preferentially located close to the interface and in the center of the channel. On the other hand, when the wall permeability is enhanced, the particle concentration becomes smoother and the particle layering is attenuated as demonstrated by the reduction in the particle concentration close to the porous interface. Overall, this indicates that the particle tendency to migrate towards the wall at high concentration is partially counteracted by the wall permeability.

\begin{figure}[t]
  \centering
  \includegraphics[width=0.45\textwidth]{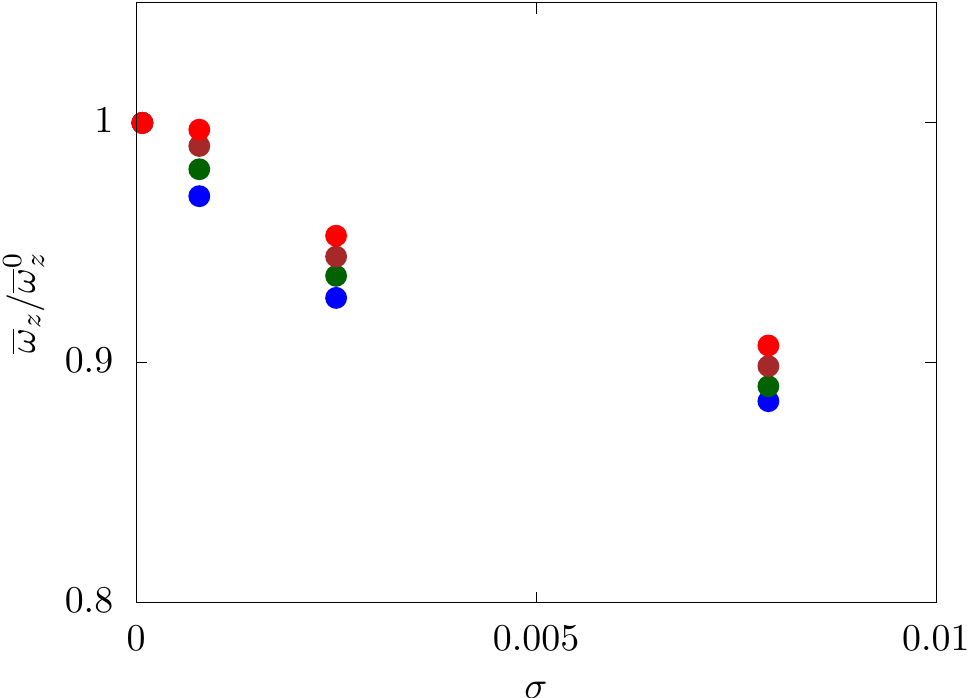}
  \caption{Spanwise component of the particle angular velocity $\overline{\omega}_z$ as a function of the wall permeability $\sigma$ for the different volume fractions considered. The angular velocity is normalized by the value obtained over rigid walls $\overline{\omega}_z^0$. The blue, green, brown and red symbols are used to distinguish $\Phi=0.06$, $0.12$, $0.24$ and $0.3$.}
  \label{fig:vort}
\end{figure}

\begin{figure*}[t]
  \centering
  \includegraphics[width=0.32\textwidth]{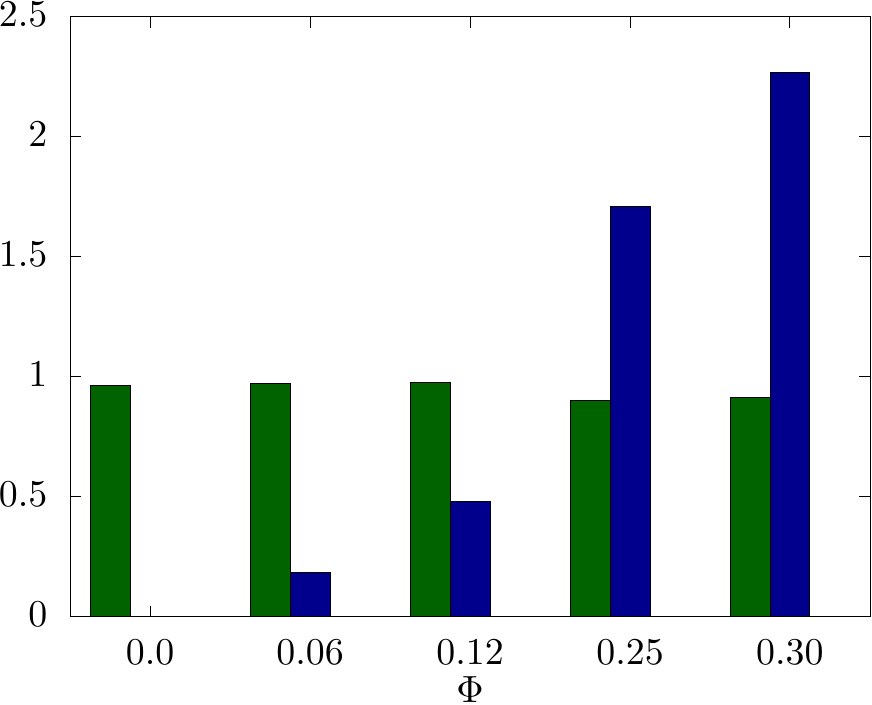}
  \includegraphics[width=0.32\textwidth]{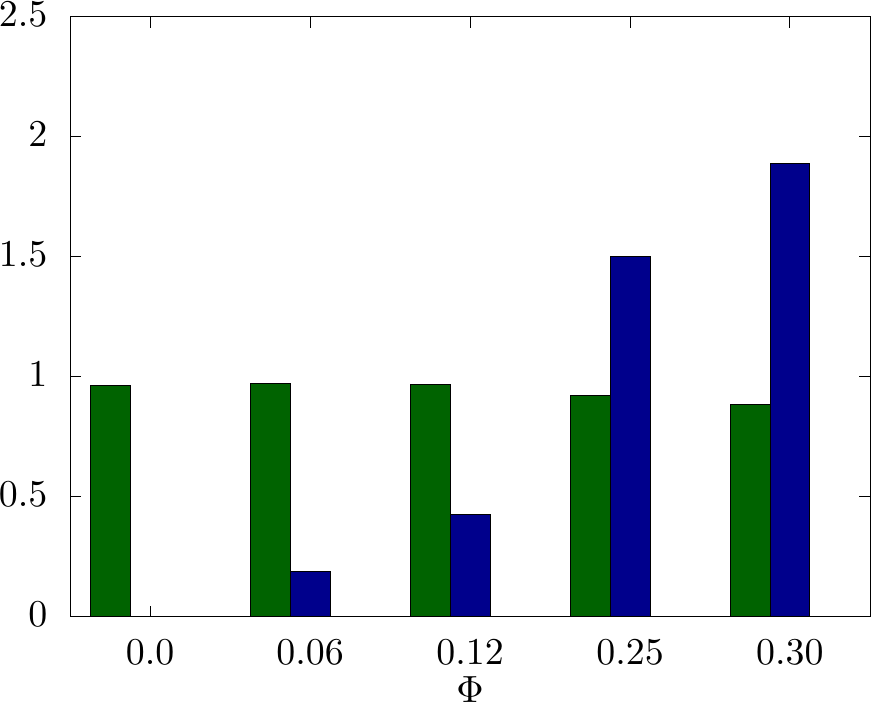}
  \includegraphics[width=0.32\textwidth]{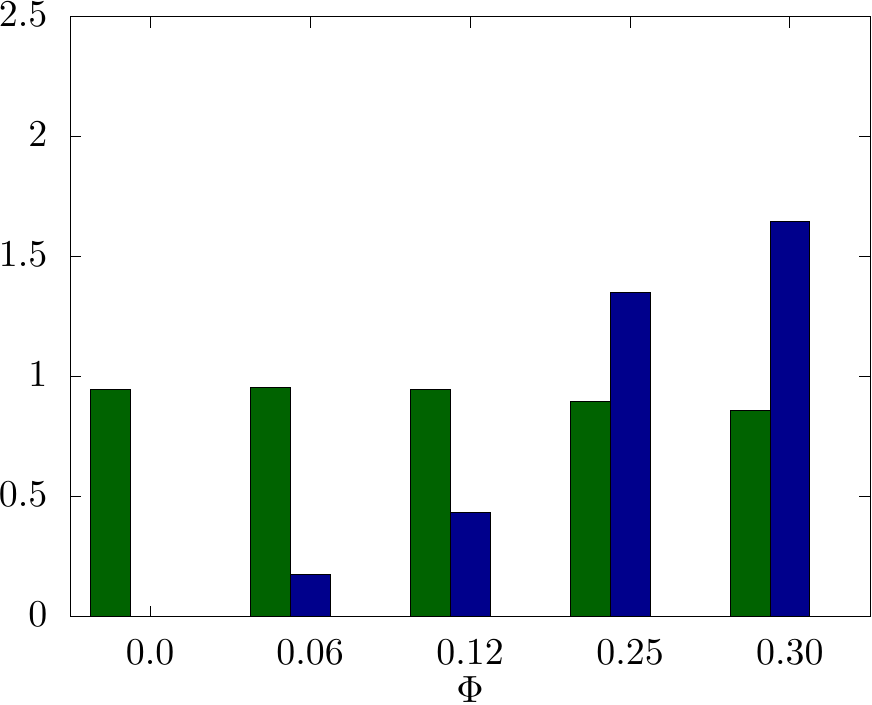}
  \caption{Histograms showing the different components of the mean shear stress balance as a function of the particle volume fraction $\Phi$ for (left) $\sigma=0.79 \times 10^{-3}$, (middle) $\sigma=2.5 \times 10^{-3}$ and (right) $\sigma=7.9 \times 10^{-3}$. The green, blue and gray colors are used to distinguish the viscous stress $\overline{\tau}_{12}^{visc}$, the particle contribution $\overline{\tau}_{12}^{part}$ and the Reynolds shear stress $\overline{\tau}_{12}^{reyn}$, with the latter hardly visible at the Reynolds number considered here.}
  \label{fig:hist}
\end{figure*}

Next, we consider the spanwise component of the mean particle angular velocity $\overline{\omega}_z$, depicted in \figrefSC{fig:vort} as a function of the wall permeability $\sigma$ and normalized by the value obtained for rigid walls. As observed in the figure, the angular velocity decreases with $\sigma$, with a maximum reduction of approximalety $10\%$ for the most permeable case considered here. The reduction of $\overline{\omega}_z$ with $\sigma$ is mainly due to the decrease in the mean shear rate across the channel caused by the wall permeability; see \figrefS{fig:vel_ref} and \ref{fig:vel}. A direct consequence of the reduced particle rotation is a lower level of fluctuations in the channel, as these are mainly generated by the particles. In summary, the analysis of the different statistical data  reveal that the presence of permeable walls decreases the mean shear across the channel, which affects the particle dynamics by reducing their rotation, in turn inducing a reduction of the velocity fluctuations due to the lower level of interaction among the suspended objects.

To gain further insight on the global suspension behavior, we examine the shear stress balance and study how the total shear stress is affected by the wall permeability and by the presence of a solid phase; in particular, we focus our attention on the purely fluid region of the domain, \ie $0<y<2h$. For a plane Couette flow, the mean total shear stress $\overline{\tau}_{12}$ can be decomposed into the sum of the viscous shear stress $\overline{\tau}_{12}^{visc} = \mu d\overline{u}/dy$, the Reynolds shear  stress $\overline{\tau}_{12}^{reyn} = -\rho \overline{u'v'}$ and the particle contribution $\overline{\tau}_{12}^{part}$ (found here by subtracting the other two components from the total shear stress), with their sum being a constant independent of the distance from the wall and equal to the total shear stress, \ie $\overline{\tau}_{12} = \overline{\tau}_{12}^{visc} + \overline{\tau}_{12}^{reyn} + \overline{\tau}_{12}^{part}$. The average value of all these different terms is reported in \figrefS{fig:hist}. The viscous shear stress remains almost constant for all the volume fractions and different wall permeabilities considered here, while the Reynolds stress contribution is very small, even not perceivable on the selected scale, due to the low Reynolds number. The particle contribution grows with the volume fraction $\Phi$ and decreases with the wall permeability $\sigma$, rapidly becoming the main contribution to the total shear stress as the volume fraction grows.

\begin{figure}[t]
  \centering
  \includegraphics[width=0.45\textwidth]{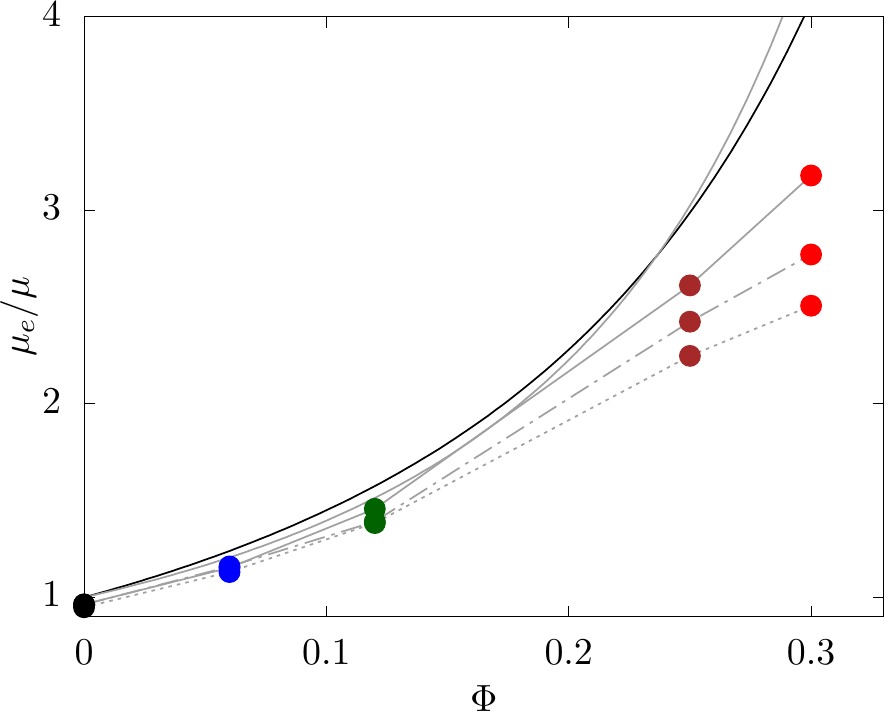}
  \includegraphics[width=0.45\textwidth]{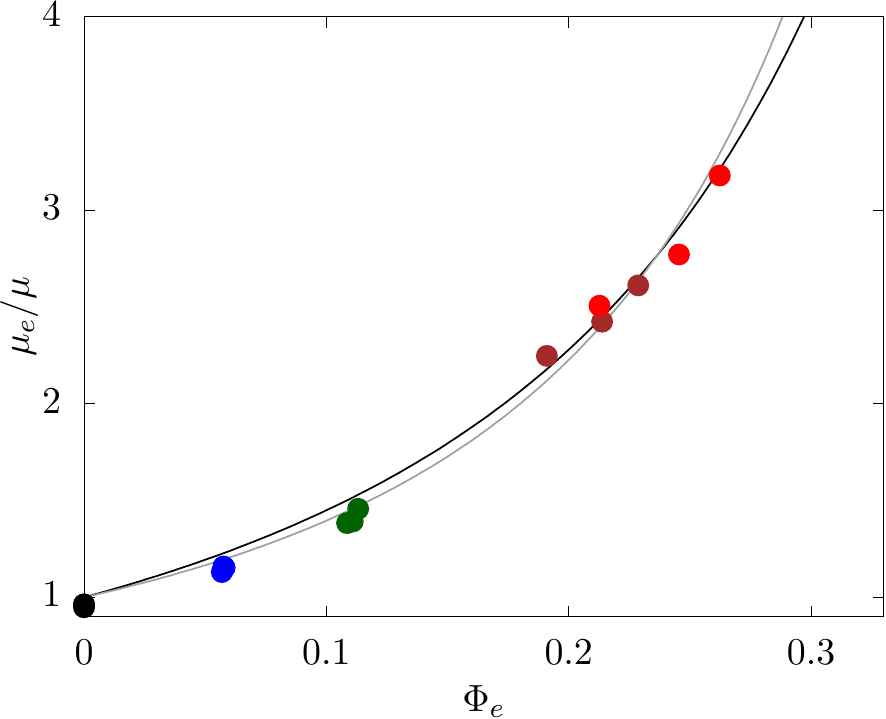}
  \caption{Effective viscosity $\mu_e$ as a function of (top) the particle volume fraction $\Phi$ and (bottom) the effective particle volume fraction $\Phi_e$ for various permeabilities $\sigma$. The black, blue, green, brown and red symbols are used to distinguish $\Phi=0$, $0.06$, $0.12$, $0.24$ and $0.3$, respectively. The black and gray solid lines are the Eilers fit \citep{eilers_1941a} and the formula by Zarraga \etal \citep{zarraga_hill_leighton-jr_2000a}.}
  \label{fig:visc}
\end{figure}
The value of the shear stress can be used to compute the so-called effective viscosity of the suspension \cite{ferrini_ercolani_de-cindio_nicodemo_nicolais_ranaudo_1979a, zarraga_hill_leighton-jr_2000a, singh_nott_2003a, kulkarni_morris_2008a, mewis_wagner_2012a, morris_2020a},  $\mu_e$, defined as $\mu_e = \overline{\tau}_{12}/ \dot{\gamma}_0$, where $\dot{\gamma}_0$ is the reference shear rate, $2U_w/2h$. The normalized effective viscosity $\mu_e/\mu$ as a function of the particle volume fraction $\Phi$ is reported in \figrefS{fig:visc}, together with the Eilers fit \citep{eilers_1941a, picano_breugem_mitra_brandt_2013a} 
\begin{equation}\nonumber
\mu_e/\mu =   \left[ 1+ 1.7 \frac{\Phi}{ \left( 1 - \Phi/0.6 \right)} \right]^2,
\end{equation}
and the formula by Zarraga \etal \citep{zarraga_hill_leighton-jr_2000a}
\begin{equation}\nonumber
\mu_e/\mu =   \frac{e^{-2.34 \Phi}}{ \left( 1 - \Phi/0.6 \right)^3},
\end{equation}
which well describe the effective viscosity of a suspension of rigid spherical particles in the reference case of impermeable walls. We observe that $\mu_e$ increases with $\Phi$ independent of the wall permeability; however, the growth rate is reduced when $\sigma$ is increased. The effect of the wall permeability becomes more pronounced as the volume fraction increases, suggesting a complex dependency of $\mu_e$ on these two parameters. We relate the decrease in the effective viscosity $\mu_e$ with the wall permeability $\sigma$ to a reduction of the wall-blocking effect due to the weakening of the no-penetration conditions; this results in a nonzero wall-normal velocity at the interface, \ie $v'_s$, as reported in \figref[bottom]{fig:wall}. This effect can be modeled as an increase in the total volume available to the particles and thus an effective reduction in the particle volume fraction $\Phi$, which we call the effective volume fraction $\Phi_e$. In particular, we compute $\Phi_e$ by increasing the total volume to include the part of the porous layer where the wall-normal velocity fluctuations $v'$ are greater than zero (more precisely where the fluctuations are larger than $0.05v'_s$). If we replot the effective viscosity now as a function of the effective volume fraction $\Phi_e$ (see bottom panel of  \figrefS{fig:visc}), we obtain a good collapse of all the data onto the experimental fits for rigid walls, which supports the idea that the reduction of $\mu_e$ can be explained in terms of an effective reduction of the suspension solid volume fraction, $\Phi$, due to the weakening of the wall effect.

\subsection{Effect of the porosity $\varepsilon$ and the porous layer thickness $h_p$}
\begin{figure}[t]
  \centering
  \includegraphics[width=0.45\textwidth]{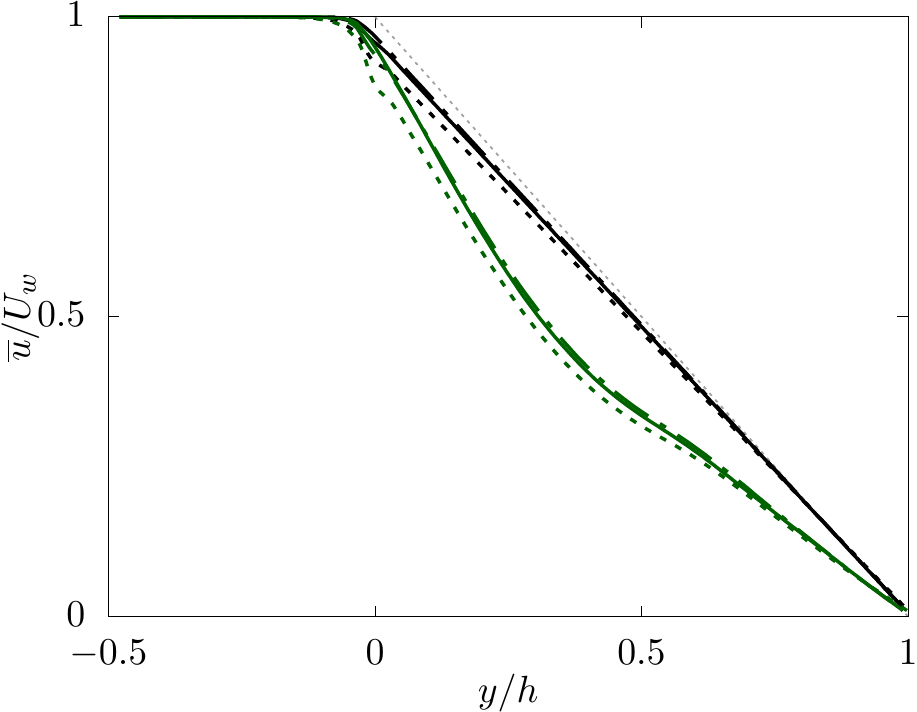}
  \includegraphics[width=0.45\textwidth]{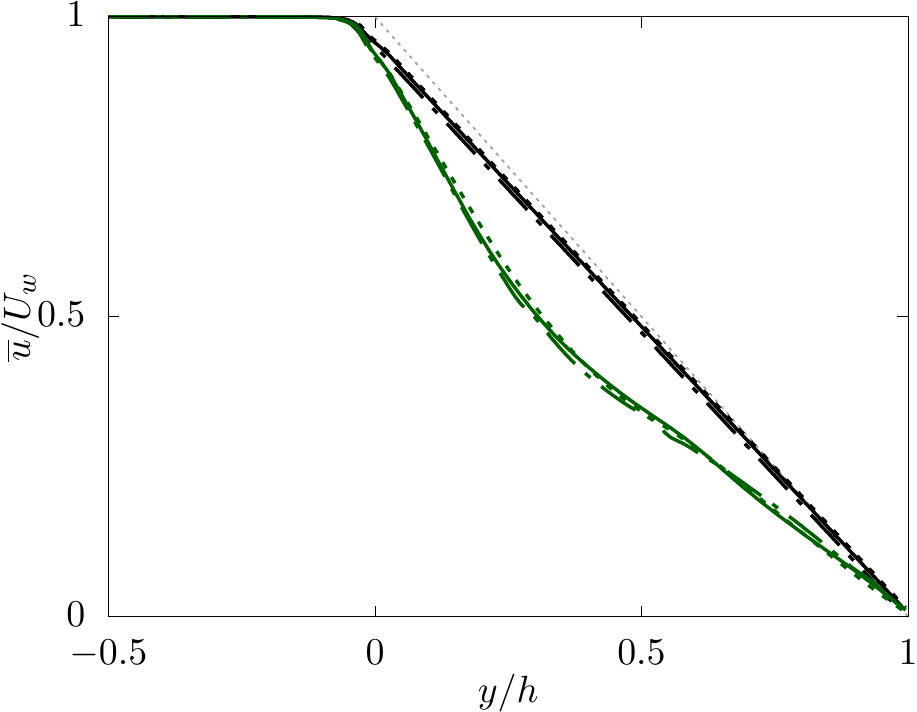}
  \caption{Mean fluid streamwise velocity component $\overline{u}$ as a function of the wall-normal distance $y$ for (top) various wall porosities $\varepsilon$ and (bottom) various thicknesses of the porous layer $h_p$. The wall permeability is fixed equal to $\sigma=7.9 \times 10^{-3}$, and two particle volume fractions are considered: $\Phi=0$ (black) and $0.12$ (green). In the figures, the dashed, solid and dashed-dotted lines are used for $\varepsilon=0.3$, $0.6$ and $0.9$ and for $h_p=0.25h$, $0.5h$ and $h$, respectively.}
  \label{fig:vel_epsh}
\end{figure}
Finally, we evaluate the effect of the other two parameters describing the porous media, \ie the porosity $\varepsilon$ and the thickness of the porous layer $h_p$.
To do so, we perform additional simulations for the single phase case $\Phi=0$ and for an intermediate volume fraction $\Phi=0.12$ both with the largest value of permeability considered in the present study $\sigma=7.9 \times 10^{-3}$; in particular, we vary the porosity in the range $\varepsilon \in \left[ 0.3,0.9\right]$ and the porous layer thickness in the range $h_p/h \in \left[ 0.25,1\right]$. The mean streamwise velocity profiles pertaining to these additional cases are reported in \figrefS{fig:vel_epsh}. We observe that for the two volume fractions considered, the variations due to these parameters are small, indicating that the variations in the permeability provide the major contribution, consistent with what was previously observed by Rosti \etal \cite{rosti_cortelezzi_quadrio_2015a}. More precisely, we find no appreciable differences when changing the porous layer thickness \cite{zhang_prosperetti_2009a, mirbod_andreopoulos_weinbaum_2009a, kang_mirbod_2019a, haffner_mirbod_2020a}, while small variations are found when changing the porosity. In particular, the two largest values of $\varepsilon$ ($0.6$ and $0.9$) provide very close results, while the smallest value of $\varepsilon$ ($0.3$) leads to a slight reduction in the mean shear rate by further increasing the slip velocity \cite{rosti_cortelezzi_quadrio_2015a}. The latter can be explained by the shear stress interface condition in \equref{eq:OTW_cond}, which prescribes an increase in the momentum jump for $\epsilon \rightarrow 0$.

\section{Conclusions} \label{sec:conclusion}
We have studied the rheology of suspensions of rigid, spherical particles in a Newtonian fluid in wall-bounded shear flow, \ie Couette flow, at a sufficiently low Reynolds 
number so that inertial effects are negligible. The part of the channel filled with particles is bounded by two rigid, homogeneous and isotropic porous layers, fixed on the moving walls and moving with the wall velocity. The problem is solved numerically using an IBM to account for the rigid suspension, while we model the presence of the porous layer by the Volume-Averaged Navier-Stokes (VANS) equations; these neglect the microscale geometry and dynamics within the porous interstices and provide a macroscopic description of the medium. The volume-averaged equations are obtained by assuming a strong separation of scales between the microscopic characteristic size of the pores and the macroscopic size of the porous medium (its thickness) and of the rigid particles, whose size is therefore large to prevent them from entering the porous medium.

We examine the rheology of the suspension by discussing how the suspension effective viscosity $\mu_e$ is affected by variations in the particle volume fraction $\Phi$ and by the level of permeability of the walls $\sigma$. We observed that $\mu_e$ is a nonlinear function of both parameters $\mu_e=\mu_e \left( \Phi, \sigma \right)$. In particular, the suspension of rigid particles has a lower viscosity in the presence of permeable walls than that measured in the case of rigid walls; this is due to the permeability of the walls, weakening the wall-blocking effect and allowing a nonzero velocity at the interface, quantified here by the slip velocity $U_s$, which grows with both $\Phi$ and $\sigma$. The rise of the slip velocity $U_s$ with the wall permeability $\sigma$ effectively reduces the mean shear rate in the domain, thus causing a reduction of the particle rotation (\ie the particle spanwise angular velocity is reduced) and a reduced interparticle interaction. The latter is the ultimate factor responsible for the reduced effective viscosity of the suspension, which we have shown to be due to a reduced particle-induced stress in the total shear stress budget.

The presence of the particles induces velocity fluctuations in the domain, which are also nonzero at the interface in the case of permeable walls. The velocity fluctuations can penetrate deeply within the porous media, where they are ultimately dissipated. We have shown that the penetration depth of the wall-normal fluctuations can be used to compute a reduced effective volume fraction $\Phi_e$, which successfully collapses all the different rheological curves $\mu_e$ vs $\Phi$ for different wall permeabilities $\sigma$ onto a single master curve, which is well approximated by the Eilers fit valid for the viscosity of a suspension flowing over rigid and impermeable walls. This suggests that the effect of the porous walls can be understood in terms of a reduced volume fraction due to the weakening of the wall-blocking effect, ultimately confirming the reduced level of particle-particle interaction. It should be noted that for single particles near a wall the properties and structure of the porous surface modify the lubrication interaction on a length scale on the order of the geometric mean of the radius and surface separation distance.  This lubrication interaction might impact the angular velocity of the particles. These are the object of our current investigations.

Moreover, this study can be seen as a step forward towards understanding the role of porous walls in the particle stress $\Sigma^p$ and the normal stress differences $N_1$ and $N_2$.  Several studies examined the particle stress in various geometries with impermeable walls \cite{gadala-maria_acrivos_1980a, acrivos_mauri_fan_1993a, zarraga_hill_leighton-jr_2000a, morris_boulay_1999a, singh_nott_2003a, sierou_brady_2002a, yurkovetsky_morris_2008a, deboeuf_gauthier_martin_yurkovetsky_morris_2009a, yeo_maxey_2010a, boyer_guazzelli_pouliquen_2011a}. In all of these studies, it was found that the particle pressure is small at low volume fraction but grows with $\phi$, reaching magnitudes of the same order as the shear stress.

Finally, we extended the use of an effective volume fraction to porous media in order to predict the suspension rheology with simple empirical fits, such as the Eilers formula, as previously done in Refs.\ \cite{picano_breugem_mitra_brandt_2013a, mueller_llewellin_mader_2010a, rosti_brandt_mitra_2018a, rosti_ardekani_brandt_2019a} for inertial effects, particle shape, deformability and wall elasticity. This scaling confirms that viscous dissipation is still the dominant mechanism at work in these flows.

\section*{Acknowledgments}
PM has been supported in part by the National Science Foundation Award No.\ 1854376 and in part by the Army Research Office Award No.\ W911NF-18-1-0356. LB acknowledges financial support from the Swedish Research Council (VR) and the INTERFACE research environment (Grant no. VR 2016-06119) and from Grant No.\ VR 2014-5001. Computer time was provided by the Swedish National Infrastructure for Computing (SNIC) and by the Scientific Computing section of Research Support Division at OIST.

\vspace{0.5cm}


\end{document}